
\input harvmac

\rightline{Bicocca-FT-00-21}
\rightline{RI-10-00}
\Title{
\rightline{hep-th/0011046}
}
{\vbox{\centerline{The Spectrum of $N=3$ String Theory on 
$AdS_3 \times G/H$}}}

\medskip
\centerline{\it Riccardo Argurio$^1$, Amit Giveon$^2$ and 
Assaf Shomer$^2$}
\vskip .3in
\centerline{{}$^1$ Dipartimento di Fisica}
\centerline{Universit\`a Milano Bicocca}
\centerline{Piazza delle Scienze 3, Milano, Italy}
\vskip .2in
\centerline{{}$^2$ Racah Institute of Physics}
\centerline{The Hebrew University}
\centerline{Jerusalem 91904, Israel}
\vskip .2in

\centerline{\tt Riccardo.Argurio@mi.infn.it, giveon@vms.huji.ac.il }

\centerline{\tt shomer@cc.huji.ac.il}

\vskip .5in

\noindent
We continue the study of string theory on $AdS_3 \times SU(3)/U(1)$ and 
$AdS_3 \times SO(5)/SO(3)$. We compute the spacetime spectrum of the  
$N=3$ supersymmetric dual CFT using worldsheet techniques. 
The spectrum of chiral primaries coincides
for the two models. Unlike $N=4$ theories, the building block of the symmetric 
product in spacetime (corresponding to a single long string) is not by itself 
in the moduli space of a symmetric product.

\vfill

\Date{11/00}

\def\journal#1&#2(#3){\unskip, \sl #1\ \bf #2 \rm(19#3) }
\def\andjournal#1&#2(#3){\sl #1~\bf #2 \rm (19#3) }

\def\frac#1#2{{#1\over#2}}

\def\half{\frac12}

\def\d{\partial}

\def\inbar{\,\vrule height1.5ex width.4pt depth0pt}
\def\IC{\relax\hbox{$\inbar\kern-.3em{\rm C}$}}
\def\IR{\relax{\rm I\kern-.18em R}}
\def\IP{\relax{\rm I\kern-.18em P}}

%
%

%
\catcode`\@=11
\def\slash#1{\mathord{\mathpalette\c@ncel{#1}}}
\overfullrule=0pt

\def\CC{{\cal C}}

\def\MM{{\cal M}}
\def\NN{{\cal N}}

\def\QQ{{\cal Q}}

\def\YY{{\cal Y}}

\def\eps{\epsilon}

\def\underrel#1\over#2{\mathrel{\mathop{\kern\z@#1}\limits_{#2}}}

\catcode`\@=12


%


\newsec{Introduction}

\lref\fms{D.\ Friedan, E.\ Martinec, and S.\ Shenker, ``Conformal
invariance, supersymmetry, and string theory," Nucl.\ Phys.\ {\bf B271}
(1986) 93.}
\lref\gk{A. Giveon and D. Kutasov, ``Little string theory in a double scaling
limit,'' JHEP {\bf 9910} (1999) 034, hep-th/9909110;
``Comments on double scaled little string theory,'' 
JHEP {\bf 0001} (2000) 023, hep-th/9911039.}
\lref\ags{R.~Argurio, A.~Giveon and A.~Shomer,
``Superstring theory on $AdS_3\times G/H$ and boundary $N = 3$ superconformal
symmetry,'' JHEP {\bf 0004} (2000) 010, hep-th/0002104.}%
\lref\gks{A.~Giveon, D.~Kutasov and N.~Seiberg,
``Comments on string theory on $AdS_3$,''
Adv.\ Theor.\ Math.\ Phys.\  {\bf 2} (1998) 733,
hep-th/9806194.}%
\lref\kuse{D.~Kutasov and N.~Seiberg,
``More comments on string theory on $AdS_3$,''
JHEP {\bf 9904} (1999) 008,
hep-th/9903219.}%
\lref\efgt{S.~Elitzur, O.~Feinerman, A.~Giveon and D.~Tsabar,
``String theory on $AdS_3 \times S^3 \times S^3 \times S^1$,''
Phys.\ Lett.\  {\bf B449} (1999) 180,
hep-th/9811245.}%
\lref\ss{A.~Schwimmer and N.~Seiberg,
``Comments on the $N=2, N=3, N=4$ superconformal algebras in two-dimensions,''
Phys.\ Lett.\  {\bf B184} (1987) 191.}%
\lref\dmvv{R.~Dijkgraaf, G.~Moore, E.~Verlinde and H.~Verlinde,
``Elliptic genera of symmetric products and second quantized strings,''
Commun.\ Math.\ Phys.\  {\bf 185} (1997) 197,
hep-th/9608096.}%
\lref\vw{C.~Vafa and E.~Witten,
``A strong coupling test of S duality,''
Nucl.\ Phys.\  {\bf B431} (1994) 3,
hep-th/9408074.}%
\lref\yis{S.~Yamaguchi, Y.~Ishimoto and K.~Sugiyama,
``$AdS_3$/CFT$_2$ correspondence and space-time $N=3$ superconformal
algebra,'' JHEP {\bf 9902} (1999) 026,
hep-th/9902079.}%
\lref\fks{J.~Fuchs, A.~Klemm and M.~G.~Schmidt,
``Orbifolds by cyclic permutations in Gepner type superstrings and in
the corresponding Calabi-Yau manifolds,''
Annals Phys.\  {\bf 214} (1992) 221.}%
\lref\klsc{A.~Klemm and M.~G.~Schmidt,
``Orbifolds by cyclic permutations of tensor product conformal field
theories,'' Phys.\ Lett.\  {\bf B245} (1990) 53.}%
\lref\orbif{L.~Dixon, J.~A.~Harvey, C.~Vafa and E.~Witten,
``Strings on orbifolds,''
Nucl.\ Phys.\  {\bf B261} (1985) 678; L.~Dixon, D.~Friedan, E.~Martinec
and S.~Shenker,
``The conformal field theory of orbifolds,''
Nucl.\ Phys.\  {\bf B282} (1987) 13.}%
\lref\mardav{F.~Larsen and E.~Martinec,
``U(1) charges and moduli in the D1-D5 system,''
JHEP {\bf 9906} (1999) 019
hep-th/9905064, J.~R.~David,
``String theory and black holes,''
hep-th/9911003.}%
\lref\ms{J.~Maldacena and A.~Strominger,
``$AdS_3$ black holes and a stringy exclusion principle,''
JHEP {\bf 9812} (1998) 005,
hep-th/9804085.}%
\lref\dijk{R.~Dijkgraaf,
``Instanton strings and hyperKaehler geometry,''
Nucl.\ Phys.\  {\bf B543} (1999) 545,
hep-th/9810210.}
\lref\deboer{J.~de Boer,
``Six-dimensional supergravity on $S^3 \times AdS_3$ and $2d$ conformal field
theory,'' Nucl.\ Phys.\  {\bf B548} (1999) 139,
hep-th/9806104.}%
\lref\kll{D.~Kutasov, F.~Larsen and R.~G.~Leigh,
``String theory in magnetic monopole backgrounds,''
Nucl.\ Phys.\  {\bf B550} (1999) 183,
hep-th/9812027.}%
\lref\dBps{J.~de Boer, A.~Pasquinucci and K.~Skenderis,
``AdS/CFT dualities involving large $2d$ $N = 4$ superconformal symmetry,''
hep-th/9904073.}%
\lref\dst{F.~Defever, S.~Schrans and K.~Thielmans,
``Moding Of Superconformal Algebras,''
Phys.\ Lett.\  {\bf B212} (1988) 467. }%
\lref\sw{N.~Seiberg and E.~Witten,
``The D1/D5 system and singular CFT,''
JHEP {\bf 9904} (1999) 017,
hep-th/9903224.}%
\lref\mo{J.~Maldacena and H.~Ooguri,
``Strings in $AdS_3$ and $SL(2,R)$ WZW model. I,''
hep-th/0001053.}%
\lref\gr{A.~Giveon and M.~Rocek,
``Supersymmetric string vacua on $AdS_3\times \NN$,''
JHEP {\bf 9904} (1999) 019,
hep-th/9904024.}%
\lref\bl{D.~Berenstein and R.~G.~Leigh,
``Spacetime supersymmetry in $AdS_3$ backgrounds,''
Phys.\ Lett.\  {\bf B458} (1999) 297,
hep-th/9904040.}%
\lref\gkp{A.~Giveon, D.~Kutasov and O.~Pelc,
``Holography for non-critical superstrings,''
JHEP {\bf 9910} (1999) 035,
hep-th/9907178.}%
\lref\ofer{O.~Feinerman,
``String theory on $AdS_3\times S^3\times S^3\times S^1$,''
M.Sc Thesis (The Hebrew University of Jerusalem).}%
\lref\malpri{J.~Maldacena, private communication.}
\lref\difr{P.~Di Francesco, P.~Mathieu, and D.~S\'en\'echal
``Conformal Field Theory,'' Springer (1997).}%
\lref\spec{R.~Argurio, A.~Giveon and A.~Shomer,
``Superstring theory on $AdS_3$ times a coset manifold,"
Proceedings of the conference Non-perturbative Quantum Effects 2000,
Paris, PRHEP-tmr2000/007, and paper to appear.}
\lref\sv{A.~Strominger and C.~Vafa,
``Microscopic Origin of the Bekenstein-Hawking Entropy,''
Phys.\ Lett.\  {\bf B379} (1996) 99,
hep-th/9601029.}
\lref\mms{J.~Maldacena, G.~Moore and A.~Strominger,
``Counting BPS black holes in toroidal type II string theory,''
hep-th/9903163.}%
\lref\sympro{R.~Argurio, A.~Giveon and A.~Shomer,
``Superstrings on $AdS_3$ and symmetric products,''
hep-th/0009242.}%
\lref\kaz{Y.~Kazama and H.~Suzuki,
``New $N=2$ superconformal field theories and superstring compactification,''
Nucl.\ Phys.\  {\bf B321} (1989) 232.}%
\lref\gko{P.~Goddard, A.~Kent and D.~Olive,
``Virasoro algebras and coset space models,''
Phys.\ Lett.\  {\bf B152} (1985) 88.}%
\lref\msm{J.~Maldacena, J.~Michelson and A.~Strominger,
``Anti-de Sitter fragmentation,''
JHEP {\bf 9902} (1999) 011, hep-th/9812073.}%

In this paper we continue the study of superstring theory on the backgrounds 
$AdS_3 \times SU(3)/U(1)$ and $AdS_3 \times SO(5)/SO(3)$.
In \ags\ we showed that superstrings propagating on these backgrounds 
give rise to a dual CFT with $N=3$ 
superconformal symmetry (or $N=1$, depending on the GSO projection).
{}Following the general result of \sympro\  the spacetime dual
CFT is expected to be in the moduli space of a symmetric product CFT
$(\MM_{6k})^p/S_p$ with $p$
the maximal number of long strings discussed in 
\refs{\sympro,\gks,\msm,\kuse,\sw,\mo}.
Here we will compute the chiral spectrum of $\MM_{6k}$ (i.e.
the untwisted sector of the dual CFT in the language of \sympro) 
using perturbative worldsheet analysis.
This will enable us to draw some conclusions about the nature of the 
spacetime CFT that were also mentioned in \sympro.
In particular, unlike other cases \refs{\gks , \efgt} which have a larger 
$(N=4)$ symmetry, the building block $\MM_{6k}$ is not
in the moduli space of any symmetric product.

\subsec{A short review of Superstrings on $AdS_3 \times G/H$}

We start by reviewing the formulation of superstrings on $AdS_3$ times
a coset CFT, following mainly the formalism of 
\refs{\gks, \kaz} and, for simplicity, treating only the holomorphic sector.
We begin by setting some notation.

The $AdS_3$ part is described by an $SL(2)$ supersymmetric WZW model that is 
constituted of the three currents $J^P$ of the $SL(2)$ affine algebra at level 
$k$, and the three fermions $\psi^P$ implied by the $N=1$ worldsheet 
supersymmetry, satisfying the following OPEs:
\eqn\slope{\eqalign{J^P(z) J^Q(w) \sim&
{{k\over 2} \eta^{PQ} \over (z-w)^2} + {i\eps^{PQR}
\eta_{RS} J^S(w) \over z-w},\cr
J^P(z) \psi^Q(w) \sim &{i\eps^{PQR} \eta_{RS} \psi^S(w)
\over z-w},\cr
\psi^P(z) \psi^Q(w) \sim &{ {k\over 2}\eta^{PQ} \over z-w},}}
where $P,Q,R,S=1,2,3$, $\eta^{PQ}=(++-)$ and $\eps^{123}=1$.

As usual in supersymmetric WZW models, the currents can be decomposed in
two pieces:
\eqn\decomp{J^P={\hat{J}}^P-{i\over k}\eta^{PQ}\eps_{QRS}\psi^R \psi^S.}
The first piece $\hat{J}^P$ constitutes an affine algebra at level $k+2$,
and has a regular OPE with the fermions $\psi^P$. We will thus refer
to $\hat{J}^P$ as the bosonic currents. The second part constitutes
an affine algebra at level $-2$.

The coset CFT part is constructed by starting from 
the $G=SU(3)$ or $SO(5)$ super affine algebra at level $k'=4k$ 
(this condition arises from criticality of the string theory 
\ags ) realized by $dimG$ affine currents $K^A$ and their fermionic
superpartners $\chi^A$, satisfying the following OPEs:
\eqn\opesu{\eqalign{K^A(z)K^B(w)\sim & {{k'\over 2}\delta^{AB} \over (z-w)^2}
+{if_{ABC} K^C(w) \over z-w}, \cr
K^A(z) \chi^B(w)\sim & {if_{ABC} \chi^C(w) \over z-w}, \cr
\chi^A(z)\chi^B(w)\sim & {{k'\over 2}\delta^{AB} \over  z-w}.}}
Here $A,B,C,D=1\dots dim G$ and $f_{ABC}$ are the structure constants of $G$.
As before, the currents split into their bosonic and fermionic parts.
The bosonic currents realize an affine algebra at level $k'-h^G$ where $h^G$
is the dual Coxeter number of $G$ which is the same in both cases
$h^{SU(3)}=h^{SO(5)}=3$.
One then proceeds to construct the $SU(3)/U(1)$ and $SO(5)/SO(3)$ coset 
SCFTs following \refs{\gko , \kaz}.

As was shown in \refs{\ags} a condition needed in order to realize an $N=3$ 
algebra is that an unbroken $SU(2)$ at level $k'$ 
must remain after modding out by $H \subset G$. Such an $SU(2)$ subgroup indeed
remains in the two coset models considered here, and we keep the indices 
$A=1,2,3$ for its generators and denote its spin by $j'$.
It was then shown in \ags\ that these two models 
indeed realize an $N=3$ superconformal symmetry in spacetime, for 
a definite GSO projection.

In the next section we will compute the chiral spectrum of the untwisted 
sector (in the sense of \sympro) of the spacetime CFTs corresponding to 
the two coset backgrounds $AdS_3 \times SU(3)/U(1)$ and $AdS_3
\times SO(5)/SO(3)$.
We therefore start with a short reminder of the chirality condition 
for the $N=3$ superalgebra and of the relation between worldsheet and 
spacetime quantities in the general context of superstrings on $AdS_3$ 
backgrounds.

\subsec{The chirality condition in the $N=3$ superconformal algebra}

The global part of the $N=3$ superconformal algebra in the NS sector
\refs{\ss,\yis} contains the following anti-commutator
\eqn\three{\{ Q_r^a, Q_s^b\} = 2 \delta^{ab} L_{r+s} +i {\eps^{ab}}_{c} (r-s)
T^c_0,} 
where $a,b,c=1,2,3$ and $r,s=\pm\half$.
We see that a unitarity bound 
\eqn\bps{2L_0\geq |T^3_0|} arises from evaluating the
norm of $Q_{-\half}^{+(-)}|h\rangle$, where $|h\rangle$ is a primary
state.
The chirality condition is the saturation of this bound $2L_0=T^3_0$, i.e.
that the spacetime weight equals half the spacetime 
$R$-charge which in our case 
arises from the unbroken $SU(2)$ spin:
\eqn\chirco{h_{ST}={1\over2}j_{ST}.}
This can, of course, occur only for the 
highest weight state inside each $SU(2)$ multiplet.
The states with $j_{ST}=-m_{ST}=2h_{ST}$ are called anti-chiral states and obey 
the opposite condition $2L_0=-T^3_0$.
Note that in the case of the $N=3$ superconformal algebra the $R$-charge is 
quantized to be in ${1\over2}{\bf Z}$ because the $U(1)$ is actually the 
$J^3$ of the $SU(2)$ $R$-symmetry.
Therefore only
the multiplicity of chiral states in each energy level is relevant.

\subsec{The relation between worldsheet and spacetime quantum numbers}

In \refs{\gks} it was shown that vertex operators transforming under the
worldsheet $SL(2)$ algebra in the representation $j$ have the following 
scaling dimension in the spacetime CFT: 
\eqn\dim{h_{ST}=j+1.}
It was also demonstrated that an affine 
algebra ${\hat G}$ in the worldsheet CFT gives rise to an affine algebra in 
the spacetime CFT such that if the worldsheet vertex operators transform under
${\hat G}$ in some representation $r^G$ then the corresponding spacetime fields
transform in the same representation $r^G$ under the spacetime affine currents.
Since we look for chiral primary states of the spacetime CFT
we have to demand two things:
\item{\it (i)} The state has to be a primary field of the spacetime CFT
and must therefore have a definite scaling dimension $h_{ST}=j_{SL(2)}+1$, 
i.e. it must transform irreducibly under the worldsheet $SL(2)$.
\item{\it (ii)} The state should have a minimal spacetime weight for every
representation of the spacetime superconformal $R$-symmetry which arises from 
the unbroken $SU(2)$.

\newsec{Computation of the chiral spectrum in spacetime}

\subsec{Vertex operators and the physicality condition - The NS sector}

We start by considering the general form of an NS vertex operator. 
The construction of such vertex operators in curved backgrounds of the form 
$AdS_3 \times G/H$ goes along similar lines to that of the flat 
case \refs{\gks , \fms}.
Here however we have to replace the plane wave zero modes with the vertex 
operators $\Phi_{j,m}$ and $U_{r,j',\QQ}$ corresponding to the zero modes on 
$AdS_3$ and $G/H$.
The vertex operators on $AdS_3$ are labeled by $j,m$ according to the 
representation of $SL(2)$ \refs{\gks}, and those on $G/H$ by the 
representation of $G$ that we denote here by $r$, by the spin $j'$ of the
unbroken $SU(2)$ and possibly by other quantum numbers of 
$G/H$ that are denoted here by $\QQ$.
The tower of string states is obtained as usual by multiplying these zero 
mode operators by a polynomial $P_N$ of scaling dimension $N$ in the fermionic
oscillators $\psi^P$, $\chi^A$ and the bosonic oscillators ${\hat{J}}^P$, 
${\hat{K}}^A$
and their derivatives (and a similar polynomial in the antiholomorphic
oscillators).
This operator has then to obey physicality conditions, usually implemented by
restricting to the BRST cohomology and performing a chiral GSO projection.
We thus write the general form of the NS vertex operator in the $(-1)$ picture,
writing for brevity only the holomorphic part:
\eqn\vert{e^{-\phi}P_N (\psi^P,\partial{\psi^P},\chi^A,\partial{\chi^A} \dots
,\hat{J}^P,\partial {\hat{J}^P},\hat{K}^A,\partial {\hat{K}^A}) \Phi_{j,m}
 U_{r,j',\QQ}.} 
The physicality condition in the case of an NS vertex operator is that it 
must be a primary field of weight $(1,1)$ in the worldsheet CFT (more
precisely, BRST invariance is equivalent to requiring that the
matter part of \vert\ is the lower component of an $N=1$ worldsheet
superfield of weight $(\half,\half)$).

Considering affine algebra primaries, which are also Virasoro
primaries,  their weight is:
\eqn\weight{\Delta_G={Q^G_{r_G} 
\over{\hat{k}_G+h^G}}~,} 
where 
$Q^G_{r_G}$  is the value of the quadratic Casimir operator of the group $G$ 
in the representation $r_G$, $\hat{k}_G$ is the bosonic level of the affine $G$,
$h^G$ is the dual Coxeter number
and we normalized to 1 the length squared of the highest root of $G$ 
(see e.g. \difr). 

The weight of a primary in a coset CFT $G/H$ is similarly obtained since
the coset CFT energy-momentum tensor is $T_{G/H} = T_G -T_H$
and thus the weight is the difference $\Delta_G -
 \Delta_H$.

Using the above\foot{Note that the total levels of the affine $G$
and $H$ are the same, because the embedding $H\subset G$ is such and 
$k'=\hat{k}_G+h^G=\hat{k}_H+h^H$, and
in both cases of interest the highest root of $H$
has the same length as the highest root of $G$, so that the normalization
in $\Delta_G$ and $\Delta_H$ does not change.},
we can write the physicality condition as:
\eqn\phys{{1\over2}+N+{-j(j+1) \over k} + {Q^G_{r_{G}}-Q^H_{r_H}  \over k'}=1.} 
Since we are interested in the spectrum of chiral primaries we have to look for 
the smallest scaling dimension in each $SU(2)$ representation. 
In a similar way to the flat-space case the smallest $N$ we can take is
$N=1/2$ because the $N=0$ vertex is tachyonic, and will not be mutually local
with the supercharges.
We therefore look for vertex operators of the general form \vert\ with $P_N$ 
being a linear combination of fermions such that the full vertex has a 
well defined $SL(2)$ and $SU(2)$ quantum numbers.
Actually, linear combinations of $\psi^{\pm,3}$ will raise or lower
the $SL(2)$ spin, while combinations of $\chi^{1,2,3}$ will do the
same for the $SU(2)$ spin (see for instance appendix A of \kll).
{}For the time being we concentrate on the first possibility, and
consider the other at the end of this subsection.

We thus pick out the smallest $SL(2)$
spin, i.e. the $j-1$ component $(\psi \Phi_j )_{j-1}$
which is \refs{\gks , \kll} (again, only the holomorphic part is shown):
\eqn\psivi{e^{-\phi}\left( \psi^3 \Phi_{j,m}- {1\over2} \psi^{-} \Phi_{j,m+1}- 
{1\over2} \psi^{+}\Phi_{j,m-1}\right) U_{r,j',\QQ},}
and has spacetime scaling dimension $h_{ST}=j$.

Taking this into account, i.e. plugging $N={1\over2}$ in \phys \ 
and using $k'=4k$ we end up with the following condition:
\eqn\fiz{j(j+1) = {1\over4} \left(Q^G_{r_{G}} - Q^H_{r_{H}} \right)~.}
We now turn to calculate the quadratic Casimir eigenvalues in a general 
representation of $SU(3)$ and $SO(5)$ in order to
compute the chiral spectrum in the $AdS_3 \times SU(3)/U(1)$ and $AdS_3
\times SO(5)/SO(3)$ cases.
To set our notation, we recall that given a representation 
$[\lambda_1,\lambda_2, \dots ,\lambda_n]$ of a Lie algebra of rank $n$,
its highest weight vector is $\mu = \lambda_1 \mu^1 + \lambda_2 \mu^2
+ \dots +\lambda_n \mu^n$, where $\mu^i$ are the fundamental weights
defined with respect to the simple roots $\alpha^i$ by 
${2 \langle\alpha^i, \mu^j\rangle \over (\alpha^i)^2} = \delta^{ij}$.
The value of the quadratic Casimir of the above representation is then 
given by
the inner product in weight-space, $Q=\langle \mu , \mu + \rho
\rangle$, where $\rho=\sum \alpha^+ $ is the sum of all positive roots.

\subsec{The $SU(3)/U(1)$ coset CFT}

The $SU(3)$ algebra is simply laced,
and in the normalization we 
use its two simple roots are: 
\eqn\rootsu{\alpha^1=\left({1\over2},{\sqrt{3}\over2}\right) ,
\quad \alpha^2=\left({1\over2},{-\sqrt{3}\over2}\right).}
Therefore the fundamental weights are \eqn\fwesu{\mu^1=\left({1\over2},
{1\over{2\sqrt{3}}}\right) ,
\quad \mu^2=\left({1\over2},{-1\over{2\sqrt{3}}}\right).}
There is one more positive root that is not simple $\alpha^3=\alpha^1+\alpha^2$
thus giving $\rho=\left(2,0 \right)$.
A general representation of $SU(3)$ is denoted by two positive integers $[r,s]$
and has a highest weight \eqn\weightt{\mu=r\mu^1+s\mu^2=\left({r+s\over2},
{r-s\over{2\sqrt{3}}}\right).}
We thus have:
\eqn\sucas{Q^{SU(3)}_{[r,s]}={(r+s)\over2}{(r+s+4)\over2}+{(r-s)^2\over{12}}.}
Now the Cartan subalgebra of $SU(3)\supset SU(2) \times U(1)$ is
composed of the $U(1)$ generator by which we mod out (denoted by $K^8$) 
and the Cartan generator $K^3$ of the $SU(2)$.
The $U(1)$ charge under this $K^8$ is the $y$ coordinate 
in weight space $(x,y)$ while the $x$ coordinate, which is always
half integer, is the $K^3$ eigenvalue.
Again, since we are looking for chiral primaries we must take the highest 
$SU(2)$ weights in a given $SU(2)$ representation. 

We now use these results to compute \fiz \ on the highest weight state of the 
$SU(3)$ representation (with weight \weightt ).
Since the charge of $H=U(1)$ is the $y$ entry in \weightt\ and since the 
quadratic Casimir of a $U(1)$ Lie algebra in a given representation 
is simply the square of the charge in that representation, we get:
\eqn\uonec{Q^{U(1)}_{[r,s]} = \left( {r-s\over{2\sqrt{3}}} \right)^2 .}
We now use \sucas\ and \uonec\
to compute \fiz\ on the highest weight, obtaining:
\eqn\sufiz{j(j+1) = {1\over4} {(r+s)\over2}{(r+s+4)\over2}, }
which is easily solved to give \eqn\jey{j={r+s\over4}.}
As explained above the $x$ coordinate in weight space $(x,y)$ is the unbroken 
$SU(2)$ charge $j'$ which serves as the $R$-symmetry in the spacetime
superconformal algebra and therefore we can write for \weightt\ 
\eqn\rsym{j_{ST}=j'={r+s\over2}.}
Comparing \rsym\ and \jey\ we conclude that  
\eqn\chirl{h_{ST}=j={r+s\over4}={1\over2}j_{ST}~,} i.e. 
these states are chiral. 

Moreover let us now prove that there are no other chiral states coming from
the same $SU(3)$ representation.
Candidate chiral states must be highest weights of an $SU(2)$
representation, and it is sufficient to consider for every value of the
$U(1)$ charge the highest weight of the largest $SU(2)$ representation.
These $SU(2)$ highest weight states are obtained from the highest weight $\mu$
of the $SU(3)$ representation labeled by $[r,s]$ by $\mu - n \alpha_1$,
with $n\leq r$, and $\mu - m \alpha_2$, with $m\leq s$.

Consider the state $\mu - n \alpha_1$. Using \weightt\ and \rootsu,
its coordinates in weight space are determined to be:
\eqn\nonchir{\mu-n\alpha_1=\left({r+s-n\over2},
{r-s-3n\over{2\sqrt{3}}}\right).}
{}From the above expression we can read off that this is the highest
weight of an $SU(2)$ representation of spin $j'={r+s-n\over2}$,
and that the $U(1)$ charge is ${r-s-3n\over{2\sqrt{3}}}$.
Let us compute the following difference:
\eqn\differ{\eqalign{j(j+1)-{j'\over 2}\left({j'\over 2}+1\right) =&
{1\over 4 }\left\{ {(r+s)\over 2}{(r+s+4)\over2}+{(r-s)^2\over 12}
- {(r-s-3n)^2\over 12}\right\} \cr & -{(r+s-n)\over4}{(r+s-n+4)\over4} \cr
 =& {n\over 4}(r+1-n) \geq {n\over 4} > 0,}}
thus proving that $h_{ST}=j>{j' \over 2}={j_{ST}\over 2}$ for a
vertex operator built from any such state, i.e. they cannot be chiral.
The same reasoning can be applied to the states $\mu-m\alpha_2$.

At this point we can prove the non-chirality 
of the spacetime fields corresponding to vertex operators with only the 
$\chi$ oscillators.
Since the fermions $\chi$ transform in the adjoint
representation of $SU(3)$ but do not carry any $SL(2)$ index the spacetime 
scaling dimension of the field corresponding to such a vertex operator will be
$h_{ST}=j+1$, and the component with largest $SU(2)$ spin in the product
$\left(\chi U\right)$ has spacetime spin $j_{ST}=j'+1$.
So using the results \jey \ and \rsym \ we can write
\eqn\contra{h_{ST}=j+1={1\over2}j'+1={(j'+1)\over2}+{1\over2}={1\over2}j_{ST}+
{1\over2} >{1\over2}j_{ST},} i.e. these states are not chiral.

\subsec{The $SO(5)/SO(3)$ coset CFT}

The $SO(5)$ algebra is not simply laced and has a length ratio of $\sqrt{2}$ 
between its two simple roots.
Normalizing the highest root to length squared 1 we can choose 
 \eqn\rootso{\alpha^1=\left({1\over2},{-1\over2}\right), 
\quad \alpha^2=\left(0,1\right).}
Therefore the fundamental weights are 
\eqn\fweso{\mu^1=\left({1\over2},0\right), \quad \mu^2=\left({1\over2},
{1\over2}\right).}
There are two more positive roots that are not simple $\alpha^3=\alpha^1+
\alpha^2=\left({1\over2},{1\over2}\right)$ and $\alpha^4=\alpha^1+
\alpha^3=\left(1,0\right)$ thus giving $\rho=\left(2,1 \right)$.
A general representation of $SO(5)$ is also denoted by two positive integers
$[r,s]$ and has a highest weight 
\eqn\weightso{\mu=r\mu^1+s\mu^2=\left({r+s
\over2},{{s\over2}}\right).}
We therefore have:
\eqn\socas{Q^{SO(5)}_{[r,s]} =
{(r+s)\over2}{(r+s+4)\over2}+{s(s+2)\over4}.}
Now, since $SO(5) \supset SO(4)=SO(3)_1 \times SO(3)_2$ the Cartan subalgebra
of $SO(5)$ is composed of the two Cartan generators $K^3_1$ and $K^3_2$ of the
two commuting $SO(3)$'s and we have to choose by which $SO(3)$ to mod out.
Choosing to mod out by the $SO(3)_2$ subgroup whose Cartan generator gives the
$y$ coordinate in weight space $(x,y)$, and again evaluating \fiz \ on the 
highest weight state of the $SO(5)$ algebra with weight \weightso \ 
for which $Q^{SO(3)_2}_{[r,s]}={s\over2}({s\over2}+1)$ we get again \sufiz.
In particular, we find also here chiral states for every spacetime 
$R$-charge $j_{ST}=j'$ coming from the NS sector of the worldsheet, built
upon an $SO(5)$ highest weight, and essentially the same reasoning as in the
previous model demonstrates that these are the only chiral states in the
NS sector.

\newsec{The Ramond sector of the worldsheet CFT}

So far we have analyzed the spectrum of chiral primary operators in the 
spacetime CFT arising from the NS sector of the worldsheet CFT.
In this section we prove that the identification of chiral primaries done so 
far is complete, i.e. that there are no new chiral primaries arising
from the Ramond sector.

A typical vertex operator in the Ramond sector can be written as:
\eqn\rvert{e^{-{\phi\over 2}} S \Phi_{j,m} U_{r,j',{\cal Q}}, }
where $S$ is a spin-field (to be discussed below) and it is assumed that
we have taken an appropriate linear combination such that the total
$SL(2)$ and unbroken $SU(2)$ spins are fixed. Again, we are considering
the lowest dimensional operator with respect to the possible excitations
as described generally in \vert.
With this taken into account, the condition for \rvert\ to be
a worldsheet primary of weight 1 is again \fiz, as for the NS vertex
operators.

The quantum numbers carried by the spin-fields $S$ were analyzed in \ags.
We can write a general spin-field as:
\eqn\spifi{S=e^{{i\over 2}(\epsilon_1 H_1+\epsilon_2 H_2 +
\epsilon_3 H_3 +\epsilon_0 \sqrt{2} H_0)}.}
The scalars in the expression above can be obtained as follows.
By bosonization, we get:
\eqn\bosonew{-\d H_1= {2\over k}\psi^1 \psi^2, \qquad
-\d H_2= {2\over k'}\chi^1 \chi^2, \qquad
-i\d H_3= {1\over k}\psi^3 \chi^3.}
Then the remaining scalar $H_0$ can be defined, for instance,
through:
\eqn\currhz{K^3=\hat{K}^3+i\d H_2 +{i\over \sqrt{2}} \d H_0.}
Alternatively, it can be constructed by bosonization of a specific combination
of the remaining fermions on $G/H$ (see \ags\ for details).

A spin-field \spifi\ has $SL(2)$ spin $j=\half $ and $SU(2)$ spin $j'=1$.
The total $SL(2)$ and $SU(2)$ spins of \rvert\ are thus
respectively $j\pm \half $ and $j'\pm 1$ or $j'$.
The candidate chiral states in the R sector can be constructed using
the same products of vertex operators used to build
the chiral states from the NS sector, and taking the combinations
with the spin-fields which lead to $SL(2)$ spin $j-\half$ and
$SU(2)$ spin $j'+1$. Indeed, we then have:
\eqn\chirr{h_{ST}=j-\half +1=j+\half={j'\over 2} +\half={j'+1\over 2}
= {j_{ST}\over 2},}
where we used \jey\ and \rsym\ which imply that $j={j'\over 2}$
for such vertex operators.

We now have to explicitly check the BRST invariance
of the vertex operators \rvert. The non-trivial part of this boils down
to checking that the OPE of the worldsheet supercurrent with \rvert\
has no $z^{-3/2}$ singular term. 

Let us see in more detail the combination \rvert\ giving rise to 
the desired spins. It is clear that if our aim is to write the
vertex operator corresponding to a chiral operator in spacetime, 
this should be the highest weight of an $SU(2)$ representation.
We will thus concentrate on the $m'=j'+1$ component of the vertex
operator with total $SU(2)$ spin $j'+1$. This will be obtained
using spin-fields which are highest weights of the spin 1 representation.
We denote them by $S^+$ and they correspond to taking $\epsilon_2=
\epsilon_0=+1$ in \spifi, as it can be determined from \currhz.

These spin-fields, labeled also by their $SL(2)$ momentum, are given by:
\eqn\spifp{S^+_\half=e^{{i\over 2}( H_1+ H_2 + \epsilon_3 H_3 +\sqrt{2} H_0)},
\qquad\qquad S^+_{-\half} =-\epsilon_3 c_1 c_3 
e^{{i\over 2}( - H_1+ H_2 - \epsilon_3 H_3 +\sqrt{2} H_0)},}
where $c_i$ are cocycle factors such that $c_i^2=1$ and $c_ic_j=-c_jc_i$.
Note that we did not yet fix the GSO projection, which will amount to
fixing the sign of $\epsilon_3$. The prefactor $-\epsilon_3$ in
$S^+_{-\half}$ is determined by requiring that the two above spin-fields
correctly transform as a spin $\half$ representation under the
$SL(2)$ currents given by:
\eqn\sltwocu{J^3=\hat{J}^3+i\d H_1, \qquad J^\pm=\hat{J}^\pm
\mp c_1c_3 e^{\pm i H_1}\left(e^{iH_3}-e^{-iH_3}\right).}
The total spin $j-\half$ combination can be written as:
\eqn\totspm{{\cal V}_{j-\half, m}= S_\half \Phi_{j,m-\half} -
S_{-\half} \Phi_{j,m+\half}.}
Thus the candidate chiral state in the R sector is given by:
\eqn\rchi{\YY=\left( e^{{i\over 2}( H_1+ H_2 + \epsilon_3 H_3 +\sqrt{2} H_0)}
\Phi_{j,m-\half}
+\epsilon_3 c_1 c_3 e^{{i\over 2}( - H_1+ H_2 - \epsilon_3 H_3 +\sqrt{2} H_0)}
\Phi_{j,m+\half}\right) U_{j'=2j, m'=2j}^{r, {\cal Q}}.}
There are three terms in the worldsheet supercurrent $G$ which give rise
to $z^{-3/2}$ singular terms in the OPE with $\YY$.
The first term is the one which comes from the terms trilinear in the
fermions, while the other two come from the terms
including respectively the bosonic $SL(2)$ and $SU(2)$ currents:
\eqn\gterms{\eqalign{G_f=& {1\over \sqrt{k}}c_3 \left[ i\partial H_1
\left(e^{iH_3}-e^{-iH_3}\right)+\half \left( i\partial H_2 +{i\over \sqrt{2}}
\partial H_0\right)\left(e^{iH_3}+e^{-iH_3}\right) \right], \cr
G_J =&  {1\over \sqrt{k}} \left[ c_1e^{iH_1}\hat{J}^- +c_1e^{-iH_1}\hat{J}^+
+c_3 \left(e^{iH_3}-e^{-iH_3}\right) \hat{J}^3\right], \cr
G_K=& {1\over 2 \sqrt{k}}c_3 \left(e^{iH_3}+e^{-iH_3}\right) \hat{K}^3.}}
We have neglected all terms which lead to less singular or regular
behavior in the OPE with $\YY$. For a derivation of the
full supercurrent $G$, see \ags.

As a warm up, we can take the OPE of $G$ with the spin-fields 
$S^+_{\pm\half}$. 
The only term contributing to the $z^{-3/2}$ singularity is $G_f$,
and the singularity is avoided taking $\epsilon_3=+1$.
This determines the BRST invariant operators giving
rise to the (global) spacetime $N=3$ supercharges, more precisely the highest
weights of the spin 1 multiplets of them.
This agrees with the result of \ags, that the $N=3$ superalgebra in spacetime
is recovered only for this GSO projection\foot{Note that there are some 
differences in the sign conventions with respect to \ags, specifically, the
signs in the bosonizations \bosonew\ here and (3.7) in \ags\ are opposite.}.

Returning to $\YY$, since $G_f$ only sees the spin-fields in $\YY$, its OPE is
easily determined to be:
\eqn\opegf{G_f(z)\YY(0) \sim z^{-{3\over 2}} {1-\epsilon_3\over 2} \YY'(0),}
where $\YY'$ is:
\eqn\yprime{\eqalign{\YY'={1\over \sqrt{k}} 
( c_3 & e^{{i\over 2}( H_1+ H_2 - \epsilon_3 H_3 
+\sqrt{2} H_0)} \Phi_{j,m-\half} \cr 
-&\epsilon_3 c_1  e^{{i\over 2}( - H_1+ H_2 + \epsilon_3 H_3 +\sqrt{2} H_0)}
\Phi_{j,m+\half} ) U_{j'=2j, m'=2j}^{r, {\cal Q}}. \cr}}
Similarly one can work out the OPEs of the other pieces of $G$, giving:
\eqn\opegc{\eqalign{G_J(z)\YY(0) \sim & z^{-{3\over 2}} \epsilon_3 (j+1) 
\YY'(0), \cr G_K(z)\YY(0) \sim & z^{-{3\over 2}} j \YY'(0).}}
The total contribution to the singularity of the OPE of $G$ with
$\YY$ is thus:
\eqn\opegtot{G(z) \YY(0) \sim z^{-{3\over 2}} {1+\epsilon_3 \over 2}
(2j+1) \YY'(0).}
This singularity is removed, and $\YY$ leads to a physical state,
only if $\epsilon_3=-1$. Since this is the opposite GSO projection to the one 
used in 
obtaining the $N=3$ algebra we conclude that these operators are projected out 
and are not part of the physical spectrum\foot{Note that performing this 
``wrong" GSO projection one ends up with a different spacetime theory with only
$N=1$ superconformal symmetry, and an additional affine $SU(2)$ algebra
that acts trivially on the supercharges \ags. Among the (now physical) 
Ramond sector operators are the lower components of the $SU(2)$ currents
with respect to this spacetime $N=1$ supersymmetry.}. 

We have thus demonstrated that there are no operators coming from
the R sector which are chiral states of the $N=3$ superconformal
spacetime theory.

\newsec{Summary and comments on the worldsheet analysis}

To summarize, we showed that every $SU(3)$ ($SO(5)$) representation gives rise
to a chiral state in spacetime, through the vertex operator \psivi\ built
using the highest weight of the representation. Moreover, we proved that these
are the only vertex operators that correspond to chiral states in spacetime.
We can read off the multiplicity of chiral states at each spacetime spin 
$j_{ST}=j'$ to be $2j'+1$ since we have such a state coming
from each $[r,s]$ $SU(3)$ ($SO(5)$) representation with ${r+s\over2}=j'$,  i.e. 
$[2j',0] ,\  [2j'-1,1] ,\  \dots ,\  [0,2j']$.
Note that this multiplicity is not the degeneracy inside a 
spin $j'$ representation. What we find are  $2j'+1$ different $SU(3) \ (SO(5))$ 
representations each leading to a chiral primary in spacetime which is 
the highest weight $(m=j')$ of an $SU(2) \subset SU(3) \ (SO(5))$ 
representation, itself with $2j'+1$ members.

The multiplicities of the chiral spectrum are the same for the two CFTs dual to 
the string backgrounds $AdS_3 \times SU(3)/U(1)$ and $AdS_3 
\times SO(5)/SO(3)$.
However one can easily get convinced that 
the non-chiral spectra are not identical. 
{}For example, in the $SO(5)/SO(3)$ model 
the state with lowest spacetime scaling dimension that is a scalar of the 
$SU(2)$ $R$-symmetry satisfies
$h_{ST}(h_{ST}+1)={1\over 8}$. Such states do not exist in the
$SU(3)/U(1)$ spacetime CFT where the neutral state of lowest dimension 
satisfies $h_{ST}(h_{ST}+1)={1 \over 4}$.
We therefore conclude that the agreement of the 
chiral spectrum suggests that the two models are sitting at two 
different points of the same moduli space.

Indeed we can identify three marginal deformations 
of the dual CFT.
The two coset models discussed here are ``geometrically rigid,'' as opposed to
\refs{\gks,\efgt} where, for instance, radii of various $U(1)$'s 
appear as NS moduli of the worldsheet sigma model.
However, we can identify three marginal deformations that come from the 
R sector of the worldsheet CFT.
We start from the three $j_{ST}=1$ chiral operators with weight $h_{ST}={1\over2}$. 
Taking the commutator of these with the spacetime supercharges which have 
the same quantum numbers we can pick out the $SU(2)$ scalar from that
product. This operator has a weight and $R$-charge obeying
$h_{ST}=1, \ j_{ST}=0$ and is thus a marginal 
deformation of the $N=3$ superconformal algebra. Furthermore it is an exactly
marginal deformation because it is the upper component of a chiral 
operator\foot{Strictly speaking, we considered in the
above paragraph only ``single particle'' marginal deformations. In this
$N=3$ set up, where the smallest chiral state has $h_{ST}={1\over 4}$,
it should be possible (if the statistics allows it)
to consider a chiral two-particle state
with $h_{ST}={1\over2}$. Then one would proceed as above to build an
exactly marginal deformation. Of course these states cannot be seen from
the perturbative worldsheet perspective.}.

A last remark is that the unitarity bound \refs{\gk,\mo}
$-\half <j< {k-1\over2}$ on the $SL(2)$ spin imposes a
bound $j'<{k'\over 4}-1$ on the $SU(2)$ spin which is generically 
stricter than the one required for unitarity of the $G/H$ CFT.

\newsec{Comments about the nature of the spacetime CFTs}

Having computed the untwisted part of the 
chiral spectrum of the spacetime CFTs dual to the 
string backgrounds 
$AdS_3 \times SU(3)/U(1)$ and $AdS_3 \times SO(5)/SO(3)$, which turns out to 
be the same, we will now draw some conclusions about the nature of these
dual CFTs.

{}Following the general result of \sympro\ we know that 
the structure of the spectrum of the spacetime dual
CFT is the same as that of a symmetric product CFT $(\MM_{6k})^p/S_p$. 
In some cases \refs{\gks , \efgt} one can go further and show that the 
building block model 
$\MM_{6k}$ itself is expected to be in the moduli space of a symmetric 
product CFT.
Specifically, for the background $AdS_3 \times S^3 \times T^4$ the 
``untwisted" 
part of the dual CFT (in the language of \sympro) was argued to be in the 
moduli space of the small $N=4$ CFT on $(T^4)^k/S_k$ (where
$k$ is the level of both the $SL(2)$ and $SU(2)$ WZW models), and for 
the background $AdS_3 \times S^3 \times S^3 \times S^1$ 
the ``untwisted" part of the dual CFT was argued to be in the 
moduli space of the large $N=4$ CFT on 
$(\MM_3)^{k'}/S_{k'}$ (where $\MM_3$ is a $c=3$ theory of one scalar and its 
four fermionic superpartners, and $k'=2k$ is the level of both 
$SU(2)$ WZW models).
This property is related to U-duality and the simple brane
picture associated with such high supersymmetry examples.
As discussed in section 6 of \sympro\ this led to the expectation that 
the full dual CFT in these two cases is also in the moduli space of
$(T^4)^{kp}/S_{kp}$ and $(\MM_3)^{k'p}/S_{k'p}$, respectively.

One might therefore ask whether a similar thing will happen in our case, i.e.
is the building block $\MM_{6k}$ also in the moduli space of a symmetric 
product CFT? If so this would mean that the full spacetime CFT will be in the
moduli space of a 
symmetric product of (an integer of order $kp$) copies of a building block with 
some fixed and small central charge.
This expectation seems to ``make sense arithmetically" 
also because in \ags\ it was shown that 
the central charge of the spacetime CFT in the two
models is $\tilde{c}={3\over2}\tilde{k}={3\over2}k'p$ where $k'$ is the level of
the affine $\hat{G}$, thus suggesting a building block CFT with central charge
${3\over2}$ and an $N=3$ superconformal symmetry. 
The unique candidate for such a building block was introduced  in \ss \ 
and is composed of a scalar $X$ compactified on a circle at the 
self dual radius, or an
$SU(2)_1$ WZW model, together with a free fermion uncharged under this 
$SU(2)$. This is the only such model since, seen as an $N=2$ minimal
model, it corresponds to the unique one with $c={3\over 2}$.
Denoting this building block theory by $\CC$ one might thus look at the
symmetric product $\CC^{k'} / S_{k'}$ (which has the right central charge 
${3\over2}k'$ and which inherits the $N=3$ superconformal symmetry in the
form of the diagonal algebra) as a candidate CFT to be in the moduli space of 
$\MM_{6k}$.

It turns out however that unlike the $N=4$ examples, 
$\MM_{6k={3\over 2}k'}$ cannot be itself
in the moduli space of some symmetric product
of a smaller building block.
The reason for that is that the chiral spectrum has
a multiplicity of $2j'+1$ chiral states for every $R$-charge $j'$, while in
any symmetric product the multiplicity of chiral states 
corresponding to ``single particle states"
at every energy (charge) level is
bounded from above by the total number of chiral operators in the building
block theory.

To see this we adapt here
the results of section 5 of \sympro\ to
show that there is a one-to-one correspondence between
chiral states in a building block theory $\MM$ (including the identity) and
chiral states in each of the $Z_N$ twisted sectors of the symmetric product 
$\MM^{k'}/S_{k'}$.\foot{Recall 
that the spectrum of the $Z_N$ twisted sectors of $\MM^{k'}/S_{k'}$ 
corresponds to operators associated with single particle states, on which we
focus in this paper.}
Moreover this correspondence is such that the chiral spectrum of the   
$Z_N$ twisted sector is a shifted version of the chiral spectrum of $\MM$
itself.
Specifically one can show \sympro\ that a chiral 
state in $\MM$ with $R$-charge $R$ and weight ${R \over 2}$ leads to 
a unique chiral state in $\MM^{k'}/S_{k'}$ 
coming from the $Z_N$ twisted sector with
weight $h^N$ and $R$-charge $R^N$ that satisfy:
\eqn\chitw{h^N=\half R +{c\over 12}(N-1)~, \qquad
R^N=R +{c\over 6}(N-1)~.}
This means that as $N$ increases by $1$ the spectrum of chiral $R$-charges is 
shifted by ${c\over6}$.
Since the step by which the weights are translated is constant,
it is straightforward to get convinced that indeed 
the multiplicity of chiral states at every $R$-charge value is
bounded from above by the total number of chiral operators in $\MM$.

One can actually compute directly the multiplicities of the chiral
operators in $\CC^{k'}/S_{k'}$. This is done by first  writing the
extended partition function of the model $\CC$, which is the product
of the extended partition function of an $SU(2)_1$ WZW model and that
of a free fermion - the Ising model.
Then using the general theory of cyclic orbifold CFTs \refs{\klsc , \fks} 
(see also appendix B in \sympro) one can identify the weight and $R$-charge 
of states in the $Z_N$ twisted sectors, and find the chiral states
for every such sector.
The result is that for every $Z_N$ sector with $N$ odd there are
exactly two chiral operators with $R$-charge $j_{ST}={N \pm 1 \over4}$.
We thus conclude that for any given $R$-charge $j_{ST}$ there are exactly 
two chiral operators that come from the $Z_N$ sectors with $N=4j_{ST} \pm 1$.
This result agrees of course with that of the arguments above.
Indeed there are exactly two chiral operators in $\CC$ which are the identity 
and the $j_{ST}={1 \over 2}$ operator $e^{{i\over{\sqrt{2} } }X }$.

\newsec{Conclusions}

Summarizing, we computed the chiral spectrum of the 
untwisted sector of the spacetime CFTs dual to the two  
string backgrounds $AdS_3 \times SU(3)/U(1)$ and $AdS_3 \times SO(5)/SO(3)$
and showed that in both cases 
for every $R$ charge $j_{ST}$ there are $2j_{ST}+1$ chiral operators 
originating from 
the NS sector on the worldsheet.
The dual CFT is expected to be $(\MM_{6k})^p/S_p$ \sympro.
Inspired by the $N=4$ cases where the 
$c=6k$ building block theory itself was argued to be in the moduli space of a 
symmetric product, we asked whether or not this is true also 
for the two $N=3$ models at hand, and showed that the answer was negative.
 
Actually, if one looks at $N=2$ cases this phenomenon where the 
building block theory itself is in the moduli space of a 
symmetric product generically fails due to simple 
algebraic considerations \refs{\sympro , \gkp}.
Consider for example the case of a background of the form 
$AdS_3 \times S^1 \times \MM_n$  where $\MM_n$ is the $N=2$ 
minimal model with central charge $c_n=3-{6 \over n}$.
On the worldsheet this is realized as the sigma model
$SL(2)_k \times U(1) \times \MM_n$.
Solving for criticality of the worldsheet theory one finds that $k=1-{1\over
n+1}$. This is generally not an integer so trivially the CFT $\MM_{6n\over n+1}$
cannot be in the moduli space of a symmetric product, even arithmetically, 
because its central charge is not linearly ``quantized."

We can thus view  the $N=3$ coset examples 
as intermediate cases between the most
symmetric $N=4$ case for which the symmetric product ansatz for $\MM_{6k}$
works, and the 
generic $N=2$ cases where such an ansatz does not even make sense.
Indeed in the $N=3$ case there was a sensible and unique candidate, and its  
spectrum did reproduce some of the features calculated in string theory,
such as the existence of a chiral operator for every 
$R$-charge $j_{ST}$, however the 
bottom line is negative.
Note also that the arguments of \refs{\dijk , \mms , \sw} 
in favor of a spacetime CFT based
on a more general symmetric product in the $N=4$ case 
rely on U-duality and the simple
brane picture of these highly symmetric cases that possibly do not apply here.
Another possibility is that at some other points in the moduli space of
the symmetric product some multiparticle states are marginally bound and 
can thus be matched with the single particle states computed from string theory.
It is more likely however that the dual CFT is simply not a symmetric product.

{}Finally we should comment that there is another model that shares the same 
spacetime symmetry structure, namely 
$AdS_3 \times (SU(2) \times SU(2) \times U(1) )/Z_2$ 
\yis\ which is a $Z_2$ orbifold of the model with large $N=4$  
superconformal symmetry \efgt.
A natural question that arises is the connection between this model and the two
models discussed in this paper,
specifically whether or not they belong to the same moduli space.
The results of this paper lead us to expect that the spacetime theory of 
this string background is not in the same moduli space as the two spacetime
CFTs discussed in this paper.
The reason is that $AdS_3 \times (SU(2) \times SU(2) \times U(1) )/Z_2$ is an
orbifold of a theory which has the property that $\MM_{6k}$ is in the moduli
space of a symmetric product, and thus it might also share this property, 
while the CFTs dual to  
$AdS_3 \times SU(3)/U(1)$ and $AdS_3 \times SO(5)/SO(3)$ where shown not to 
possess this property.

\bigskip
\noindent{\bf Acknowledgments:}
We are happy to thank S.\ Elitzur for many valuable discussions
and G.~Sarkissian for pointing out important references.
R.A. thanks the Service de Physique Th\'eorique at the
Universit\'e Libre de Bruxelles and especially F.~Englert
and M.~Henneaux for their warm hospitality and for partial
support while this work was being completed.
A.G. thanks the Theory Division at CERN for its hospitality.
This work is supported in part by the BSF -- American-Israel Bi-National
Science Foundation -- and by the Israel
Academy of Sciences and Humanities -- Centers of Excellence Program.

\listrefs

\end